\documentstyle[prl,aps,epsfig,times]{revtex}
\begin{document}
\draft
\twocolumn[\hsize\textwidth\columnwidth\hsize\csname @twocolumnfalse\endcsname
 \title{
Finite-Size Studies on the 
$SO(5)$ Symmetry of the Hubbard Model 
 }
 \author{Stefan Meixner, Werner Hanke}
\address{Institut f\"{u}r Theoretische Physik, Am Hubland,
D-97074 W\"{u}rzburg, Federal Republic of Germany}
 \author{Eugene Demler and Shou-Cheng Zhang}
 \address{
 Department of Physics, Stanford University, Stanford, CA 94305
 }
 
\date{\today}

\maketitle
 \begin{abstract}
We present numerical evidence for the approximate $SO(5)$ symmetry
of the Hubbard model on a $10$ site cluster. Various dynamic
correlation functions involving the $\pi$ operators, the generators 
of the $SO(5)$ algebra, are studied using exact diagonalisation,
and are found to possess sharp collective peaks. Our numerical results
also lend support on the interpretation of the recent resonant neutron
scattering peaks in the $YBCO$ superconductors in terms of the 
Goldstone modes of the spontaneously broken $SO(5)$ symmetry.

 \end{abstract}
 \pacs{PACS numbers: 7.10.Bg, 71.55.Jv}
]

\tighten
\narrowtext

Although the single band Hubbard model has been extensively 
studied in recent years in connection with high-$T_c$ 
superconductivity, its low-energy content has so far eluded
both analytical and numerical investigations. 
While being deceptively simple,
this model may have many possible competing ground states, and it
has proved to be very difficult to organize the low-energy degrees of
freedom. Recently, a new analytic approach based on a symmetry 
principle has been suggested. It was noticed that the Hubbard model
enjoys an approximate $SO(5)$ symmetry, unifying antiferromagnetism (AF)
with $d$-wave superconductivity (SC)\cite{so5}. This symmetry principle
gives a simple description of the transition from an AF ground state
to a SC ground state as the chemical potential is varied, and it
gives a unified treatment of the low-energy collective degrees
of freedom of the Hubbard model. 

Motivated by its potential importance to the high-$T_c$ problem,
we undertake a numerical finite size study to test this approximate
symmetry in the single band Hubbard model. The central object of our 
study is the so-called $\pi$ operator defined as follows,
\begin{eqnarray}
\pi_{s/d}^\dagger= \sum_{\bf{k}} (\cos k_x \pm \cos k_y) \;
      c_{{\bf k} + {\bf Q},\uparrow}^\dagger 
      c_{-{\bf k},\uparrow}^\dagger
      \  \ {\bf Q}=(\pi,\pi)
\label{pi}
\end{eqnarray}
This operator carries spin 1, charge 2 and total momentum $(\pi,\pi)$.
Charge conjugation and spin lowering operation gives five other
similar operators. Here the subscript $\alpha = s,d$ refers to the internal
$s$-wave or $d$-wave symmetry of this composite operator. 
The $\pi$ operators were first introduced
by Demler and Zhang\cite{demler} to explain the resonant neutron 
scattering peaks in the $YBCO$ superconductors\cite{neutron}, and are
constructed by following the analogy with the $\eta$
operators considered by Yang\cite{yang}. 
More recently, it was found\cite{so5} that together with the total
spin and charge operators, the six $\pi$ operators form the generators
of the $SO(5)$ algebra, and furthermore, they rotate AF
and SC order parameters into each other.
In reference \cite{demler} and \cite{so5}, it was argued that the
$\pi_d$ operators are approximate eigen-operators of the Hubbard 
Hamiltonian. This implies that the $SO(5)$ symmetry
is an approximate symmetry of the Hubbard model\cite{so5}, and one
can use it
to constrain the form of the low-energy effective Hamiltonian.

In this work, we present exact numerical diagonalisation studies
of the dynamic correlation functions involving the 
$\pi$ operators and the AF and SC order parameters. We 
verify that the $\pi$ operators are approximate eigen-operators
of the Hubbard model, and show that the properties of the various mixed 
correlation functions are consistent with the anticipated pattern of
the $SO(5)$ symmetry breaking. Our exact numerical calculation is
performed on a $\sqrt{10} \times \sqrt{10}$ site Hubbard cluster, 
using standard exact diagonalisation (ED) methods based on the Lanczos
algorithm\cite{ed_tech}. Mixed dynamical correlation functions are 
calculated utilizing a technique previously used to obtain anomalous
Green's functions\cite{ed_mixed}.

As a check to our numerical algorithm and the general framework, we
first performed our calculations on the correlation functions involving
the $\eta$ operator\cite{yang}, its mixed correlation functions with
the charge-density-wave (CDW) order parameter\cite{so4} and the 
dynamic auto-correlation function of the CDW order parameter for
the negative $U$ Hubbard model. The relationship between the 
$\eta$ and the $\pi$ operators is explained in reference 
\cite{so5}. Since the $\eta$ operator is a 
exact eigen-operator and the negative $U$ Hubbard model is known to
have a $s$-wave SC ground state away from half-filling, various 
exact results can be derived for these correlation functions\cite{so4}. 
Our numerical results are consistent with these exact results. However,
due to the space limitations, we will not present these calculations
here.

Let us first consider the auto-correlation function of the $\pi$
operators, defined as follows,
\begin{eqnarray}
     \pi^{+}_{\alpha}(\omega)= -\frac{1}{\pi} \Im
      \langle \Psi_0^{N}|\pi_{\alpha}
      \frac{1}{\omega - H+E_0^{N+2}+ i\eta}
        \pi_{\alpha}^\dagger|\Psi_0^N\rangle.
\label{pi-correlation}
\end{eqnarray}
Here $H$ is the standard Hubbard Hamiltonian, $|\Psi_0^N\rangle$
its ground state with $N$ electrons and $E_0^{N}$ the corresponding
ground state energy. $\Im$ takes the imaginary part of a
correlation function. Throughout this paper, we measure the energy
of a spectral function from the ground state energy of the intermediate
state. Since $\pi_{s/d}^\dagger$ changes particle number by two, 
the natural energy scale of the intermediate states is the ground 
state energy of a $N+2$ electron system.

\begin{figure}
\begin{center}
\epsfig{file=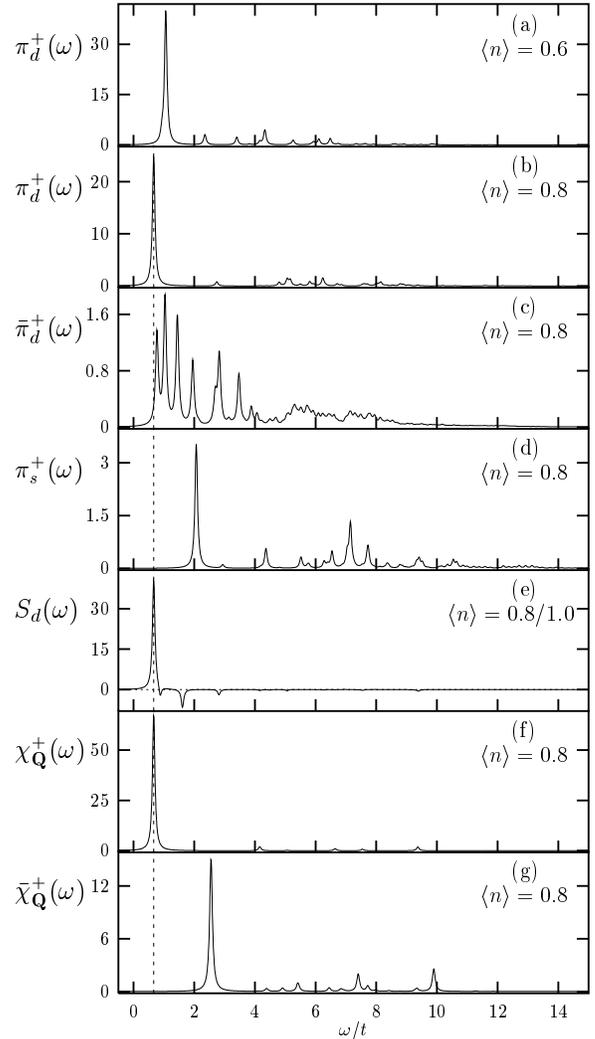,width=7.50cm}
\end{center}
\vspace*{-0.20cm}
\caption{Dynamic correlation functions of the $\sqrt{10} \times \sqrt{10}$
   Hubbard model with $U=4t$: (a) $\pi_d^+(\omega)$ spectrum at 
   $\langle n \rangle = 0.6$, (b) $\pi_d^+(\omega)$ spectrum at
   $\langle n \rangle = 0.8$, (c) bubble approximation to $\pi_d^+(\omega)$
   as shown in (b), (d) $\pi_s^+(\omega)$ at $\langle n \rangle = 0.8$, (e)
   off-diagonal correlation function $S_d(\omega)$ at 
   $\langle n \rangle = (0.8/1.0)$,
   as defined in the text, (f) spin correlation function 
   $\chi_{\bf Q}(\omega)$ and (g) bubble approximation 
   $\bar{\chi}_{\bf Q}(\omega)$.}
\label{f01}
\end{figure}

\begin{figure}
\begin{center}
\epsfig{file=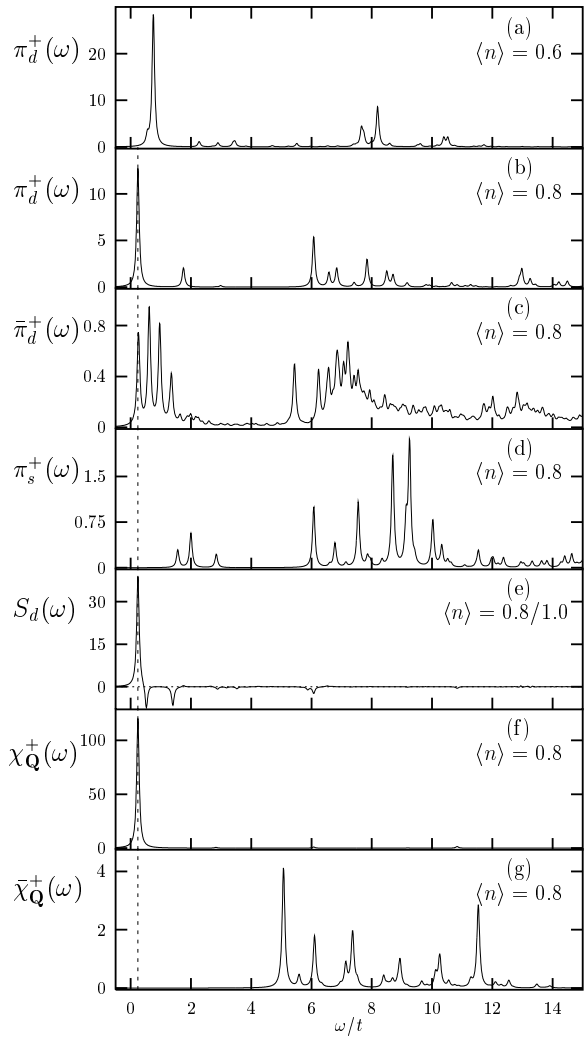,width=7.50cm}
\end{center}
\vspace*{-0.20cm}
\caption{Same as in Fig. \ref{f01}, but here for $U=8t$.}
\label{f02}
\end{figure}

Figures \ref{f01}(a), \ref{f01}(b), \ref{f02}(a) and \ref{f02}(b) 
plot the $\pi^{+}_d(\omega)$ correlation function 
for $U=4t$ and $U=8t$, with electron densities $\langle n \rangle = 0.6$
and $0.8$ respectively. We see that these dynamic correlation functions
are dominated by a sharp peak, well separated from a higher energy
continuum. If $\pi^\dagger_d$ was an exact eigen-operator of the
Hubbard Hamiltonian, its dynamical auto-correlation function should
consists of a single peak. Here we see that it is only an approximate
eigen-operator in the sense that there is also a high-energy continuum in 
addition to the resonance peak. The ratio of the spectral weight under the
peak to the total spectral weight is found to be $0.63, 0.68, 0.46$
and $0.31$ in figs \ref{f01}(a) to \ref{f02}(b). The separation between
the peak and the continuum,
and the large relative spectral weight of the peak demonstrate that the
$\pi^\dagger_d$ operator is an eigen-operator of the Hubbard model to a good
degree of approximation. 
The energies of these peaks are $1.06t$, $0.66t$, $0.55t$
and $0.24t$, respectively, and are low compared to $U$. This can easily be
understood from the fact that the $\pi$ operators are spin triplet
operators, therefore, the bare $U$ interaction should not enter its
energy scale.

In reference \cite{demler}, the dynamic auto-correlation function of the
$\pi_d$ operators was calculated within the $t$-matrix approximation in
the particle-particle channel. Within this approximation, the dynamic
auto-correlation function consists of a single peak, with a spectral
weight proportional to $1-n$, {\it i.e.} the hole density, and an energy
of $J(1-n)/2-2\mu$, where $J\approx 4 t^2/U$ is the AF
exchange constant and $\mu$ is the chemical potential measured from
half-filling. Strictly speaking, the $t$-matrix
approximation should be valid only in the low electron density limit.
Here we see that the qualitative features of the $t$-matrix calculation
are still valid near half-filling. Our exact numerical calculations 
show that {\it the $\pi^{+}_d(\omega)$ correlation function is indeed 
dominated by a single peak whose spectral weight is proportional 
to the hole density $1-n$. Furthermore, the energy of the $\pi$ resonance
peak is also proportional to the hole density, with a scale 
comparable to $J$.}

The $\pi$ resonance is a composite particle made out of two electrons. 
Near half-filling, it is very difficult for a single electron or 
hole to propagate coherently. In fact it is not clear if there is
a quasi-particle peak in the one particle Green's function at all.
In view of this fact, it is rather surprising that a two electron
Green's function $\pi^{+}_d(\omega)$ has a coherent peak while the 
individual electrons propagate incoherently. To demonstrate that 
the $\pi$ peak is a genuine collective behavior, we plot in figures
\ref{f01}(c) and \ref{f02}(c) the bubble approximation to the
$\pi^{+}_d(\omega)$ correlation function. This approximation to
$\pi^{+}_d(\omega)$ consists of a particle-particle bubble with fully
dressed one particle Green's function of the $\sqrt{10} \times \sqrt{10}$ site
cluster inside the bubble. It fully takes into
account the single-particle dressing effects, only the vertex correction
is neglected. From these figures, we see that the bubble approximation
yields a broad spectral distribution, without any identifiable resonance
peak. The height of the broad spectral distribution is one
order of magnitude less than the height of the 
$\pi$ peak found in the full calculation. 
{\it This calculation demonstrates
unambiguously that the collective behavior, or in other words, the
vertex correction, is responsible for the existence of the $\pi$ resonance.}

Figures \ref{f01}(d) and \ref{f02}(d) give results for $\pi^{+}_s(\omega)$
at $U=4t$ and $U=8t$, respectively, at density $n=0.8$. For $U=8t$, the
spectrum is broadly distributed and no resonance peak can be identified.
For $U=4t$, although there is an isolated spectral feature, it is at a
much higher energy ($\approx 2t$) 
compared to the $\pi^{+}_d(\omega)$ counterpart,
and the height of the peak is almost one order of magnitude less. 
We interpret these results as evidence that in contrast to 
$\pi^\dagger_d$,  
$\pi^\dagger_s$ is not an approximate eigen-operator
of the Hubbard model near half-filling. This result is consistent with
the conclusion of the $t$-matrix calculation\cite{demler}. The crucial
difference between the $\pi^{+}_d(\omega)$ and $\pi^{+}_s(\omega)$ 
correlation function has recently been used by Zhang\cite{so5} to
argue that {\it there is an approximate symmetry 
between AF and $d$-wave SC,
but no symmetry between AF and extended $s$-wave SC near half-filling.}

Originally, the $\pi$ operators were introduced\cite{demler}
in connection with the
resonant neutron scattering peaks recently found in the $YBCO$ 
superconductors\cite{neutron}. Below the superconducting transition
temperature $T_c$, a resonance peak appears in the polarized neutron
scattering amplitude with momentum transfer $(\pi,\pi)$, and a 
resonance energy
of $25meV$, $33meV$ and $41meV$, depending on doping. 
Demler and Zhang\cite{demler} identified this experimental feature
with the $\pi$ resonance, and showed that the particle-particle
$\pi$ resonance can be mixed into the particle-hole channel below
$T_c$, and therefore appear in the neutron scattering cross-section.
A number of other theoretical papers\cite{other} explain the 
experimental feature in terms of a particle-hole threshold behavior
near the superconducting gap or possible excitonic states inside the
gap. More recently, the $\pi$ resonance has been interpreted more
broadly by Zhang\cite{so5} as the Goldstone boson associated with
the spontaneous breaking of the $SO(5)$ symmetry below $T_c$.

In this work, we test the ideas of references \cite{demler} and \cite{so5}
by studying the mixed correlation functions involving the 
spin-density-wave order parameter
\begin{eqnarray}
S_{\bf Q}^{+}= \sum_{\bf{k}}
      c_{{\bf k} + {\bf Q},\uparrow}^\dagger c_{{\bf k},\downarrow}\ \, \ \
      S_{\bf Q}^{-}={S_{\bf Q}^{+}}^\dagger
\label{sdw}
\end{eqnarray}
and the $\pi_d$ operator, defined by
\begin{eqnarray}
     S_d(\omega)= -\frac{1}{\pi} \Im 
      \langle \Psi_0^{N-2}| 
      (\pi_d\frac{1}{\omega - (H-E_0^{N})+ i\eta}S_{\bf Q}^+ \nonumber \\
      -
       S_{\bf Q}^+ \frac{1}{\omega - (H-E_0^{N-2})+ i\eta} \pi_d)
      |\Psi_0^N\rangle
\label{mixed}
\end{eqnarray}
and the dynamical spin correlation function itself, defined by
\begin{eqnarray}
     \chi^{+}_{\bf Q}(\omega)= -\frac{1}{\pi} \Im
      \langle \Psi_0^{N}|S_{\bf Q}^{-}
      \frac{1}{\omega - (H-E_0^N)+ i\eta}
         S_{\bf Q}^{+}|\Psi_0^N\rangle
\label{chi}
\end{eqnarray}
{\it The correlation function $S_d(\omega)$ satisfies an important 
exact sum rule}
\begin{eqnarray}
  \int_{-\infty}^\infty d\omega S_d(\omega) = 
   2 \ \langle \Psi_0^{N-2}| \Delta_d  |\Psi_0^N\rangle
\label{sumrule}
\end{eqnarray}
where $\Delta_d=\sum_{\bf{k}} (\cos k_x - \cos k_y) 
      c_{{\bf k},\uparrow}
      c_{-{\bf k},\downarrow}$
is the $d$-wave superconducting order parameter. 
This sum rule follows from the fact that the $\pi_d$ operator is a
$SO(5)$ symmetry generator which rotates the AF order parameter into
the $d$-wave SC order parameter\cite{so5}. Mixed correlation
functions like $S_d(\omega)$, which involve a symmetry generator and
an order parameter, are commonly used to prove the Goldstone theorem
on the Goldstone bosons associated with 
spontaneous symmetry breaking\cite{goldstone}. 

In figures \ref{f01}(e) and \ref{f02}(e), $S_d(\omega)$ is plotted
for $U=4t$ and $U=8t$, for electron density between $n=0.8$ and $n=1$.
We see that the spectrum is dominated by a single peak with relative
weight $0.77$ and $0.72$ respectively,
located at the same energy as the $\pi$ resonance. 
We also verified that the sum rule (\ref{sumrule}) is satisfied
by explicitly computing the $d$-wave order parameter defined on the
right hand side. Therefore, our
result shows that {\it the 10 site Hubbard cluster at densities 
between $n=0.8$ and $n=1$ possesses
considerable $d$-wave SC fluctuations, and furthermore, 
there is indeed a Goldstone pole associated with this spontaneous 
symmetry breaking, consistent with the $SO(5)$ theory\cite{so5}.}
The finite energy of the Goldstone pole results from the fact that the
$SO(5)$ is not only broken spontaneously by the SC order parameter,
but also broken explicitly by the chemical potential\cite{so5}.
(Sometimes, a Goldstone boson associated with a symmetry which is
broken both explicitly and spontaneously is called a pseudo Goldstone
boson. The pion in particle physics is such an example.)
We also calculated $S_d(\omega)$ between the densities $n=0.6$ and $n=0.8$
and found it to be nearly zero. Similarly, the mixed correlation 
function $S_s(\omega)$ involving the $\pi_s$ and $S_{\bf Q}^{+}$
operators vanishes for all densities. This result is consistent with
other numerical evidence for $d$-wave pairing fluctuations in the 
Hubbard model\cite{d_wave}.

Finally we show our calculation for the spin correlation function
$\chi^{+}_{\bf Q}(\omega)$ 
in figures \ref{f01}(f) and \ref{f02}(f), for $U=4t$ and $U=8t$ 
respectively at density
$n=1$. We see that {\it 
there is a sharp resonance feature at the same energy
as the $\pi$ resonance.}
The fact that all three correlation functions $\pi^{+}_d(\omega)$, 
$S_d(\omega)$ and $\chi^{+}_{\bf Q}(\omega)$ have resonance peaks 
at the same energy clearly demonstrates that the peak in the spin 
correlation function
$\chi^{+}_{\bf Q}(\omega)$ is due to a particle-particle intermediate
state $\pi^\dagger_d |\Psi_0^{N-2}\rangle$. Because of the finiteness
of the mixed correlation function $S_d(\omega)$ and the right-hand-side
of equation (\ref{sumrule}), a particle-particle excitation at density
$n=0.8$ makes a finite contribution to the spin correlation function
at $n=1$. This feature confirms the argument of reference\cite{demler},
which explains the experimental resonant neutron scattering cross section
in terms of the collective particle-particle $\pi$ resonance.

To exhibit the collective nature of the resonance peak in 
$\chi^{+}_{\bf Q}(\omega)$, and to contrast it with single-particle
behavior, we plot in figures \ref{f01}(g) and \ref{f02}(g) 
the bubble approximation to $\chi^{+}_{\bf Q}(\omega)$
on the $\sqrt{10} \times \sqrt{10}$ site cluster.
In contrast to the full spin correlation function, the bubble 
approximation does not posses any low-energy collective poles.
These results show that it is necessary to include the vertex corrections
and the mixing into the particle-particle channels
to calculate $\chi^{+}_{\bf Q}(\omega)$, in order to exhibit its resonance
peak. These calculations have been performed recently and will be 
published elsewhere\cite{future}.

In conclusion, we have numerically verified one of the most fundamental
assumption of the $SO(5)$ theory of the Hubbard model\cite{so5}, 
which asserts
that the $SO(5)$ symmetry generators, the $\pi_d$ operators,
are approximate eigen-operators
of the Hubbard model both near and away from half-filling. Their dynamic
auto-correlation function shows well separated resonance peaks with large
relative spectral weight and low-energy. This behavior is intrinsically
collective, and can not be reproduced by any calculations which neglects
vertex corrections. In contrast to the $\pi_d$ operators, the $\pi_s$
operators do not show well separated peaks near half-filling. This 
result shows that there is only an approximate $SO(5)$ symmetry between
the AF and the $d$-wave SC order parameters, no symmetry between the AF
and the $s$-wave SC order parameters. Close to half-filling, between the
densities $n=0.8$ and $n=1$, the 10 site Hubbard cluster has finite
$d$-wave superconducting fluctuations, which lead to a non-vanishing
mixed correlation function involving the $SO(5)$ symmetry generator
and the AF order parameter. Because of the finiteness of this mixed
correlation function, there is a contribution to the dynamic spin 
correlation function from the particle-particle $\pi$ resonance. These
observations are consistent with the proposed explanation of the 
resonant neutron scattering peaks in the $YBCO$ 
superconductors\cite{demler}.

We would like to acknowledge useful discussions with Prof. D. Scalapino
and P. Hedegard who urged us to contrast the collective behavior of the
$\pi$ resonance with
the single-particle bubble approximation. 
This work is supported by FORSUPRA II, BMBF (05 605 WWA 6), ERB
CHRXCT940438 and in part by the NSF under grant numbers DMR-9400372 
and DMR-9522915.

\end{document}